\documentclass[12pt]{article}
\begin{document}
\title{The Enigma of Gravitation}
\author{B.G. Sidharth\\
International Institute for Applicable Mathematics \& Information Sciences\\
Hyderabad (India) \& Udine (Italy)\\
B.M. Birla Science Centre, Adarsh Nagar, Hyderabad - 500 063
(India)}
\date{}
\maketitle
\begin{abstract}
Even after nearly a century, it has not been possible to unify
gravitation with other fundamental forces. We argue that this is
because gravitation differs fundamentally from the others, and give
a different Planck scale formulation.
\end{abstract}
\vspace{5 mm}
\begin{flushleft}
\end{flushleft}
\section{Planck Oscillators}
Some years ago \cite{ijmpa1}, we explored some intriguing aspects of
gravitation at the micro and macro scales. We now propose to tie up
a few remaining loose ends. At the same time, this will give us some
insight into the nature of gravitation itself and why it has defied
unification with other interactions for nearly a century. For this,
our starting point is an array of $n$ Planck scale particles. As
discussed in detail elsewhere, such an array would in general be
described by \cite{ng}
\begin{equation}
l  = \sqrt{n \Delta x^2}\label{4De1d}
\end{equation}
\begin{equation}
ka^2 \equiv k \Delta x^2 = \frac{1}{2}  k_B T\label{4De2d}
\end{equation}
where $k_B$ is the Boltzmann constant, $T$ the temperature, $r$ the
extent  and $k$ is the analogues of the spring constant given by
\begin{equation}
\omega_0^2 = \frac{k}{m}\label{4De3d}
\end{equation}
\begin{equation}
\omega = \left(\frac{k}{m}a^2\right)^{\frac{1}{2}} \frac{1}{r} =
\omega_0 \frac{a}{r}\label{4De4d}
\end{equation}
We now identify the particles with \index{Planck}Planck
\index{mass}masses and set $\Delta x \equiv a = l_P$, the
\index{Planck}Planck length. It may be immediately observed that use
of (\ref{4De3d}) and (\ref{4De2d}) gives $k_B T \sim m_P c^2$, which
ofcourse agrees with the temperature of a \index{black hole}black
hole of \index{Planck}Planck \index{mass}mass. Indeed, Rosen
\cite{rosen} had shown that a \index{Planck}Planck \index{mass}mass
particle at the \index{Planck scale}Planck scale can be considered
to be a \index{Universe}Universe in itself with a Schwarzchild
radius equalling the Planck length.\\
Whence the mass of the array is given by
\begin{equation}
m = m_P/\sqrt{n}\label{A}
\end{equation}
while we have,
\begin{equation}
l = \sqrt{n} l_P, \, \tau = \sqrt{n} \tau_P,\label{3e31}
\end{equation}
$$l^2_P = \frac{\hbar}{m_P} \tau_P$$
In the above $m_P \sim 10^{-5}gms , l_P \sim 10^{-33}cm$ and $\tau_P
\sim 10^{-42} sec$, the original Planck scale as defined by Max
Planck himself. We would like the above array to represent a typical
elementary particle. Then we can characterize the number $n$
precisely. For this we use in (\ref{A}) and (\ref{3e31})
\begin{equation}
l_P = \frac{2`Gm_P}{c^2}\label{B}
\end{equation}
which expresses the well known fact that the Planck length is the
Schwarzchild radius of a Planck mass black hole, following Rosen.
This gives
\begin{equation}
n = \frac{lc^2}{Gm} \sim 10^{40}\label{C}
\end{equation}
where $l$ and $m$ in the above relations are the Compton wavelength
and mass of a typical elementary particle and are respectively $\sim
10^{-12}cms$ and $10^{-25}gms$ respectively.\\
Before coming to an interpretation of these results we use the well
known result alluded to that the individual minimal oscillators are
black holes or mini Universes as shown by Rosen \cite{rosen}. So
using the Beckenstein temperature formula for these primordial black
holes \cite{ruffinizang}, that is
$$kT = \frac{\hbar c^3}{8\pi Gm}$$
we can show that
\begin{equation}
Gm^2 \sim \hbar c\label{4e4}
\end{equation}
We can easily verify that (\ref{4e4}) leads to the value $m = m_P
\sim 10^{-5}gms$. In deducing (\ref{4e4}) we have used the typical
expressions for the frequency as the inverse of the time - the
Compton time in this case and similarly the expression for the
Compton length. However it must be reiterated that no specific
values
for $l$ or $m$ were considered in the deduction of (\ref{4e4}).\\
We now make two interesting comments. Cercignani and co-workers have
shown \cite{cer1,cer2} that when the gravitational energy becomes of
the order of the electromagnetic energy in the case of the Zero
Point oscillators, that is
\begin{equation}
\frac{G\hbar^2 \omega^3}{c^5} \sim \hbar \omega\label{4e5}
\end{equation}
then this defines a threshold frequency $\omega_{max}$ above which
the oscillations become chaotic. In other words, for meaningful
physics we require that
$$\omega \leq \omega_{max}.$$
where $\omega_{max}$ is given by (\ref{4e5}). Secondly as we can see
from the parallel but unrelated theory of phonons \cite{huang,reif},
which are also bosonic oscillators, we deduce a maximal frequency
given by
\begin{equation}
\omega^2_{max} = \frac{c^2}{l^2}\label{4e6}
\end{equation}
In (\ref{4e6}) $c$ is, in the particular case of phonons, the
velocity of propagation, that is the velocity of sound, whereas in
our case this velocity is that of light. Frequencies greater than
$\omega_{max}$ in (\ref{4e6}) are again meaningless. We can easily
verify that using (\ref{4e5}) in (\ref{4e6}) gives back (\ref{4e4}).
As $\hbar c = 137e^2,$ in a Large Number sense, (\ref{4e4}) can also
be written as,
$$Gm^2_P \sim e^2$$
That is, (\ref{4e4}) expresses the known fact that at the Planck
scale, electromagnetism equals gravitation in terms of strength.
However on using equation (\ref{A}) we get the empirically well
known electromagnetism-gravitation relation (Cf.eg.(\ref{e14})).\\
In other words, gravitation shows up as the residual energy from the
formation of the particles in the universe via Planck scales particles.
Furthermore, the above considerations mimic a degenerate Bose gas and the pressure is given by
(Cf.ref.\cite{huang}),
$$p = \alpha k T > 0$$
where $\alpha > 0$ is a suitable multiplier. This means that the force is always
attractive, indeed as is true for gravitation.\\
The scenario which emerges is the following. Analogous to Prigogine
cosmology \cite{prigogine,tryon}, from the dark energy background,
in a phase transition Planck scale particles are suddenly created.
These then condense into the longer lived elementary particles by
the above process of forming arrays. But the energy at the Planck
scales manifests itself as gravitation, thereafter.\\
We will further discuss this in the next section.
\section{Discussion}
Equation (\ref{C}) can also be written as
\begin{equation}
\frac{Gm}{lc^2} \sim \sqrt{N}\label{D}
\end{equation}
where $N \sim 10^{80}$ is the Dirac Large Number, viz., the number
of particles in the universe. There are two remarkable features of
(\ref{C}) or (\ref{D}) to be noted. The first is that it was deduced
as a consequence in the author's 1997 cosmological model \cite{uof}.
In this case, particles are created fluctuationally from the
background dark energy. The model predicted a dark energy driven
accelerating universe with a small cosmological constant. It may be
recalled that at that time the prevailing paradigm was exactly
opposite -- that of a dark matter constrained decelerating universe.
As is now well known, shortly thereafter this new dark energy driven
accelerating universe with a small cosmological constant was
confirmed conclusively through the observations of distant
supernovae. It may be mentioned that the model also deduced other
inexplicable relations like the Weinberg formula that relates the
microphysical constants with a large scale parameter like the Hubble
Constant:
\begin{equation} m \approx
\left(\frac{H\hbar^2}{Gc}\right)^{\frac{1}{3}}\label{3e6}
\end{equation}
While (\ref{3e6}) has been loosely explained away as an accidental
coincidence Weinberg \cite{weinberggravcos} himself emphasized that
the mysterious relation is in fact unexplained. To quote him, "In
contrast (this) relates a single cosmological parameter (the Hubble
Constant) to the fundamental constants $\hbar , G, c$ and $m$ and is
so far unexplained."\\
The other feature is that (\ref{D}) like (\ref{3e6}) expresses a
single large scale parameter viz., the number of particles in the
universe or the Hubble constant in terms of purely microphysical parameters.\\
As we saw the scenario is similar to the Prigogine cosmology in
which out of what Prigogine called the Quantum Vacuum, or what today
we may call Dark Energy background, Planck scale or Planck mass are
created in a phase transition, very similar to the formation of
Benard cells \cite{prig}. The energy at the Planck scale, given by
(\ref{4e5}) then gets distributed in the universe -- amongst all the
particles, as the Planck particles form these various elementary
particles according to equations (\ref{4De1d}) to (\ref{3e31}). This
is brought out by the fact that equation (\ref{D}) can also be
written as the well known Eddington formula:
\begin{equation}
Gm^2/e^2 \sim \frac{1}{\sqrt{N}}\label{e14}
\end{equation}
which was believed to be another ad hoc coincidence unrelated to
(\ref{3e6}). Equation (\ref{e14}) shows how the gravitational force
over the cosmos is weak compared to the electromagnetic force. In
other words the initial "gravitational energy" on the formation of
the Planck scale particles, that is (\ref{4e4}) is distributed
amongst the various particles of the universe \cite{bgs}. From this
point of view while $l,m,c$ etc. are indeed microphysical constants
as Dirac characterized them, $G$ is not. It is related to the Large
Scale cosmos through the Dirac Number $N$ of particles in the
universe. This would also explain the Weinberg puzzle: In this case
in equation (\ref{3e6}), there are the large scale parameters namely
$G$ and $H$ on right side of the equation.\\
Once we recognize this, we can easily see that unlike what was
thought previously, the Weinberg formula (\ref{3e6}) is in fact the
same as the Dirac formula (\ref{e14}). To see this, we use in
(\ref{3e6}) two well known relations from cosmology
(Cf.eg.\cite{weinberggravcos}), viz.,
$$R \sim \frac{GM}{c^2} \, \mbox{and} \, M = Nm$$
where $R$ is the radius of the universe $\sim 10^{28}cm$, $M$ its
mass $\sim 10^{55}gm$ and $m$ is as before the mass of a typical
elementary particle. Then (\ref{3e6}) will reduce to (\ref{e14}).
Thus, there is only one relation -- (\ref{3e6}) or (\ref{e14}), and
they express the fact that rather than being a microphysical
parameter, $G$ rather than representing a fundamental interaction is
related to the large scale cosmos via either of these equations.\\
It must be observed that this conclusion resembles that of Sakharov
\cite{sakharov}, for whom Gravitation was a secondary force like
elasticity.

\end{document}